# An Integrated Aerial Telecommunications Network that Supports Emergency Traffic

Laurent Reynaud[†], Tinku Rasheed[‡] and Sithamparanathan Kandeepan[‡§]
[†]Orange Labs, 2 Avenue Pierre Marzin, 22307 Lannion, France
[‡]CREATE-NET, via alla Cascata 56D, 38123 Trento, Italy
[§]School of Electrical and Computer Engineering, RMIT University, Melbourne, Australia
Email: laurent.reynaud@orange-ftgroup.com, tinku.rasheed@create-net.org, kandeepan@ieee.org

*Abstract*— **This paper outlines how an aerial telecommunications network can optimally meet the stringent needs of emergency relief and recovery operations. We propose a novel architecture, made of an integrated and highly dynamic multi-purpose aerial telecommunications infrastructure that can be contextually extended with fast-deploying high or low altitude platforms. In particular, we analyze the interest and challenges of adapting core concepts from substitution networks and controlled mobility mechanisms, so that a base network can be seamlessly augmented, both in terms of capacity and functions. We give an estimation of the emergency traffic supported by the lower altitude platforms in an example scenario and discuss the challenges posed by this architecture, notably in terms of disaster resilience and ability to efficiently provide sustained first responder communications.**

## I. INTRODUCTION

In the aftermath of a disaster, whether natural or man-made, it is of the utmost importance that rescue teams and other potentially life-saving response operations can rely on an efficient emergency communications system. Unfortunately, terrestrial networks may be partially or totally compromised, at least during the early stages of the post-disaster period. Similarly, alternative communications systems may prove less than effective. As an illustration, the report of the Select Bipartisan Committee to investigate the response to Hurricane Katrina [10] clearly concluded that massive network failures and severe interoperability issues with the remaining communications systems greatly hindered rescue effort, situational awareness and command and control operations. For the support of reliable emergency communications, recovery solutions relying on satellite communications have long been investigated through multiple research activities including the European projects Real-Time Emergency Management via Satellite (REMSAT) [3] and Emergency Management by Satellite Communications (EMERGSAT) [3].

In this context, aircraft-based telecommunications, which we investigate through this paper, have also gained momentum in the recent years, with many types of aerial platforms seen as a possible extension or alternative to traditional equipments from terrestrial and satellite wireless networks. Those platforms, either aerostats or aerodynes, evolve at various altitudes and are named Low Altitude Platforms (LAPs) [4] or High Altitude Platforms (HAPs) [1], [2], whether they fly in the lower troposphere or in the stratosphere. Depending on multiple factors including their altitude, available power for the payload, the type of antenna and radio technology, an aerial platform can cover areas of up to 300 km radius. Moreover, many aerial vehicles are able to keep a quasi-stationary position. Direct Line Of Sight (LOS) conditions with terrestrial equipments are thus often met. As a consequence, many use cases that take benefit of aerial platforms have been recently envisioned [1], [2], including emergency communications. Due to their payload flexibility, relatively fast deployment times and low altitude (hence allowing reduced radio propagation delays), HAPs and LAPs are increasingly seen as a viable solution to support relief services. They can quickly restore communications in the affected zones without mandatory support of the terrestrial infrastructure, which may be at least partially inoperable. Moreover, the aerial platforms can be effectively deployed together with the existing satellite-terrestrial recovery infrastructure, to provide speedy deployment able to cover larger areas and to maintain effective communication links supporting adequate capacities and bandwidth demands of the terrestrial first responders.

However, it is observed that relief scenarios based on aerial platforms lack a global architecture that can federate and support all the requirements of the multiple actors involved in emergency relief and recovery operations. On the opposite, each service or resource requirement related with an emergency role (e.g. search and rescue personnel also known as first responders, command and control operations and critical infrastructure repair teams) is currently addressed independently, in a standalone fashion.

In this paper, we present an integrated Aerial Telecommunications Network (ATN) architecture for emergency communications, constituting an integrated and highly dynamic multi-purpose aerial telecommunications infrastructure that can be contextually extended with fast-deploying aerial platforms. We select a set of typical emergency services needed by rescuers, and outline the subsequent impacts in term of required capacity on the considered network. Bandwidth requirements are estimated for each type of traffic. We then argue that such an integrated emergency communications platform is inevitable given the

This work was partially supported by the ANR RESCUE project, grant ANR-10-VERS-003 of the French Agence Nationale de la Recherche.

predictability of different disasters and the flexibility provided with fast and augmented critical infrastructure restoration support that can meet the variability of the estimated traffic requirements. We also analyze the interest and challenges of adapting core concepts from substitution networks and controlled mobility mechanisms, so that a base network can be seamlessly augmented in terms of coverage, capacity and functions.

## II. RELATED WORK

Several studies identified the benefits of aerial platforms with the objective of supporting wireless communications in emergency situations. In these works, targeted airborne platforms are not always designed for higher elevations. LAPs in particular are not only seen as a fast and convenient way to experiment telecommunications payloads with relatively inexpensive aerial vehicles, but can also address actual scenario requirements where limited coverage, due to low altitude, is acceptable. The authors in [4] and [9] sought to evaluate several LAP vehicles for relief support in several regions of Indonesia particularly exposed to a large number of potential threats, including droughts, floods, landslides, earthquakes and volcanic eruptions. In this context, a simple tethered balloon, flying at a maximum altitude of 440 m, can carry an IEEE 802.11a/g payload and cover an area of about 72 $km^2$. Naturally, in these cases, the covered area and maximum system throughput are relatively limited and only LOS communications are supported. Other radio technologies were also investigated, such as in [7]. This work, outlining a multi-hop architecture using relaying features specified in the IEEE 802.16j standard, is not specific to HAPs and may also be considered for LAPs. Still, no performance evaluation was presented and the study thus remains speculative in this regard. In contrast, a practical approach was used in [6] where emergency communications for electric power systems were investigated. It was shown that, compared to satellite systems, HAPs could favorably meet the stringent needs of power grid security in terms of both bandwidth and especially delay.

However, in a more general case, the support of emergency communications is investigated at higher altitudes. In [3], the use of satellites to recover communications in disaster-affected areas is investigated, and several deployment scenarios that combine the features of low and geostationary earth orbiting satellites and HAPs are proposed. Although the multiple use-cases clarify the advantages of integrated aerial infrastructures in disaster relief situations, no insight is given as to when and how the considered infrastructures are set up. The author of [5] investigated how first-responder and emergency answering services such as 911 or 112 could be supported by HAPs. One of the major considerations in this paper was to propose a monolithic HAP payload architecture that would enable interoperability with most representative first-responder technologies. Also, as cellular networks constitute an important communications asset for the various rescue and relief teams, the support of several technologies may be required. As a result, the specified payload consists of multiple access technologies with adapted antennas, Voice over IP (VoIP) and radio gateways, a full-featured IP core and several adapted services. It is thus unrealistic to consider that this payload architecture could be deployed according to its full specifications on a currently available HAP airplane or airship, without severe capacity limitations for each of the supported radio technologies.

## III. A FLEXIBLE AERIAL TELECOMMUNICATIONS NETWORK

In our architecture, an ATN is a collection of aircrafts able to communicate with other nodes through adapted payloads. Compared to existing network topologies which include airborne platforms [1], [2], [3], [5], we defined a multi-level hierarchical aerial topology. These levels, even if it is the case in this paper, are not necessarily related to different ranges of altitude. Instead, they are primarily intended to identify several sets of airborne platforms with common key properties such as mobility patterns, payload type, platform capacity and supported radio technologies. As an application of this general principle, Fig. 1 illustrates the desired network architecture for disaster relief operations, with two distinct levels:

• The higher level of aerial platforms consists of one or several quasi-stationary vehicles, integrated to satellite and terrestrial networks via high-capacity links. Naturally, the availability of terrestrial backhaul links also depends on the extent of damage caused by the considered disaster. However, remote access to ground stations which are located out of the damage zone should be facilitated by the support of high capacity Inter-Platform Links (IPL) and multi-hop relay mechanisms. These nodes may already be in place prior to the disaster, provided a commercial aerial network is already operated on the zone or the disaster can be anticipated to some extent. Moreover, they are assumed to be evolving at high altitudes, in order to benefit from HAPs key properties such as large terrestrial coverage, contained average wind intensity and favorable radio propagation conditions between both terrestrial and satellite segments.

• The lower level of aerial platforms is made of all the nodes that are set up and deployed for specific recovery missions. These platforms may have different mobility patterns, flight durations, payload types and communication links. As depicted by Fig. 1, the communications in this heterogeneous environment are facilitated by the support of IPL with neighboring same-level platforms, but also with higher level platforms. Note that IPL from different levels do not necessarily use the same radio technologies and the same types of antennas. In fact, higher level IPL are expected to use directive antennas and Free Space Optics (FSO), because of the relative stability of the vehicles and the favorable radio propagation conditions. However, due to the more dynamic mobility patterns and design constraints related to the lower level vehicles (including cost and weight), sectoral and omnidirectional antennas and other radio technologies are generally preferred for higher level IPL.

As a result, the ATN can be hierarchically partitioned into clusters [14], as illustrated by Fig. 2. This topological representation offers many advantages. In particular, an adapted large scale routing protocol could inherently differentiate the types of links (backhaul links and IPL) by using routing schemes adapted to each partitioned zone of the network.

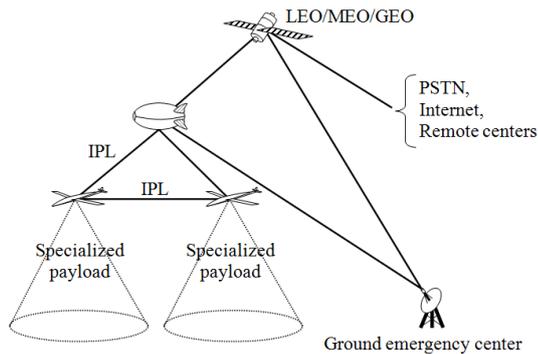

Figure 1. A 2-level aerial network adapted to disaster relief operations. Aerial vehicles are interconnected through IPL.

Furthermore, our ATN architecture is also designed to use the concept of controlled mobility [8], as Fig. 3 illustrates. In essence, a base network, experiencing troubles such as communication link instability or congestion, may be temporarily supplemented with additional wireless dirigible nodes. Moreover, an integrated ATN which federates multiple types of emergency recovery equipment is complex and costly to deploy, with a high demand in term of ground staff dedicated to the network operations [2]. Controlled mobility is intended to alleviate this demand with self organization mechanisms that take benefit of the unique mobility capabilities of aerial platforms. This way, emergency personnel may be partially relieved of some remote topology control and vehicle guidance tasks and be allowed to concentrate on their respective core rescue and recovery missions. The envisioned way to enforce controlled mobility within the integrated ATN is manifold. Yet, two approaches in particular can be outlined:

• The base network may be made of LAPs (each node following a predetermined mobility pattern) and the terrestrial infrastructure equipment left undamaged by the disaster. In this scenario, controlled mobility is applied to the HAPs, which are moved so that they can provide maximum coverage and bandwidth to terrestrial and LAP equipments.

• This base network may instead consist of the HAPs, with quasi-stationary positions, plus, again, undamaged or repaired terrestrial equipment. Controlled mobility is here applied to the LAPs. LAP default mobility patterns will thus be altered so that they can extend the base network's capacities where it is best needed. In the rest of this paper, we consider and discuss this approach.

One of the major expected benefits of the controlled mobility and other discussed ATN mechanisms is the ability to meet applicative demands in a given area with a relatively low number of nodes. In the next section, we seek to illustrate the corresponding ideal case, where a minimum number of LAPs can optimally support a variable emergency traffic demand.

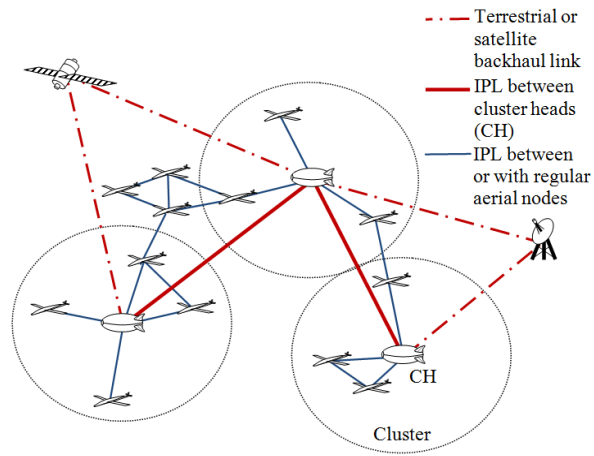

Figure 2. Hierarchical representation of the multi-level ATN topology.

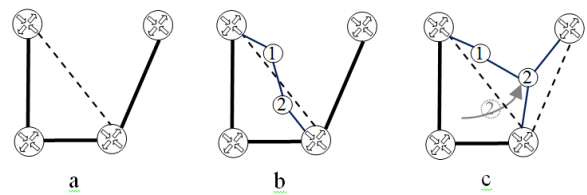

Figure 3. The concept of controlled mobility. In case a, a communication link (represented by dotted lines) cannot meet the traffic demand. In case b, mobile nodes are introduced and placed so that an additional multi-hop route is established between the nodes with the mentioned link. If another link experiences troubles, as represented in case c, the mobile nodes can be moved accordingly.

## IV. EMERGENCY TRAFFIC REQUIREMENTS

Out of the various services used by the emergency response teams, we chose to estimate the global voice emergency and medical care video traffics. In order to perform this estimation, we contextualized the emergency traffic with parameters consistent with the 2005 Hurricane Katrina disaster. Among those parameters are the affected area (233,000 km$^2$ [9]), the extent and duration of the base mobile communication network outage [5] and the estimation of the population of the first responders and related traffic [5], [10]. Note that we took the example of the Hurricane Katrina disaster because this large-scale event is well-documented, with in particular abundant literature about the composition and actions of the different response teams and about the coordination problems (partly due to inadequate communication systems) experienced by rescuers.

### A. Voice emergency traffic estimation

We define the voice emergency traffic as the total number of emergency voice calls passed by the public and first responders (e.g. rescuers, army, emergency medical care, nonmedical care and public safety teams). We base our estimation on the related traffic pattern estimated in [5] for the 20 days following Hurricane Katrina landfall, and similarly perform a normalization of the daily traffic corresponding to a daylight rescue. For each day of the corresponding period, we calculate the busy hour traffic and take this conservative value as the traffic which the ATN must support. These values are plotted in Fig. 4. Furthermore, as our considered ATN architecture is intended to achieve voice communications interoperability through a VoIP system that supports adapted gateways, we set a series of plausible hypotheses about the used radio technologies and VoIP codecs and configuration. According to [9], 78 VoIP calls can be simultaneously established with an acceptable quality from a LAP evolving at an altitude of 440 m and an achievable capacity of 54 Mb/s anywhere in a covered zone of 47.39 km$^2$. In these conditions, with an 802.11g communication interface, an Adaptive Multi-Rate (AMR) codec at a rate of 12.2 kb/s and a sample period of 20 ms, a Mean Opinion Score (MOS) of 3.8 can be achieved. Hence, using these parameters for our estimation, we calculated and represented in Fig. 4 the minimum number of LAPs required to support the corresponding traffic. A maximum of 475 concurrent voice sessions supported by 7 LAPs is attained on day 15.

*B. Video emergency traffic estimation*

In our estimation, the video emergency traffic is emitted solely by the medical assistance teams, and more precisely, in the context of Hurricane Katrina, by the Disaster Medical Assistance Teams (DMATs). In the USA, DMATs, which belong to the National Disaster Medical Systems (NDMS), are teams of physicians, nurses, technicians, and support staff. Their main mission is to provide emergency care, perform triage and resuscitation, stabilize and prepare for evacuation. At the time of Hurricane Katrina landfall, there were 45 DMATs spread out across the USA. When deployed on a disaster zone, DMATs are typically made of 35 members [11]. According to [10], there were 9 DMATs initially pre-positioned before Katrina landfall. Eventually, all the DMATs were progressively deployed in response to the disaster [12]. Our estimation reflects the ratio of prepositioned medical care units and the progressive activation of supplementary teams. Also, we took the following assumptions:

• Each deployed DMAT is made of 35 members, including 4 physicians. Those physicians work in pairs, and perform alternating 12-hour shifts. We thus consider that, at any time, there are two active physicians per DMAT. At any time, the number of video streams per DMAT is equal to the number of active physicians, i.e. 2.

• 9 DMATs are active at day 1, and this number of active DMATs progressively reaches 45 at day 15. Moreover, when deployed, we assume that each DMAT is fully operational for the remaining analyzed period.

• Also, we consider that with video sessions based on the H.264 codec with a bit rate of 384 kb/s, most medical video teleconsultation services can be supported with an acceptable quality [13]. With the same configuration in terms of altitude, ground coverage and communication interfaces as used when estimating voice traffic, each LAP can support 18 simultaneous H.264 video sessions at the given bit rate.

With these hypotheses, we estimated and represented in Fig. 5 the number of active DMATs for the observed 20-day period. The subsequent number of required concurrent video sessions is easily deduced, and allows estimating the minimum number of LAPs (also represented in Fig. 5) able to support the corresponding traffic.

*C. Aggregate emergency traffic estimation*

While the required number of LAPs for aggregate voice and video traffic generally depends on whether the LAPs are dedicated to one or both traffic types, in our example however, the same maximum value of 12 LAPs is attained in both cases, at days 15, 16 and 18. Furthermore, it shall be noted that this estimation relies on the ability to serve every applicative demand with a relatively low LAP coverage. For instance, in our example, the maximum covered LAP area without overlap is 568.68 km$^2$, and covers less than 0.25% of the overall disaster area. In the context of Hurricane Katrina, this low proportion is not unrealistic, considering the very low population density in a large part of the affected area, and the large number of rescue teams in a few key locations [10]. Nevertheless, some rescue teams may be mobile and the emergency traffic may vary with time and location. As a result, the sustained performance of the considered architecture strongly relies on the ability to take advantage of the HAP larger coverage and to enforce efficient ATN mechanisms, including controlled mobility.

V. MAIN CHALLENGES FOR FUTURE INTEGRATED ATN

One of the prominent challenges affecting the development of integrated ATN's is the technological maturity of the platforms themselves [2]. The airship-based HAPs and unmanned UAVs, both based on fuel and solar energy are at different levels of development. They are governed not only by technical and manufacturing constraints but also by both political and economic factors.

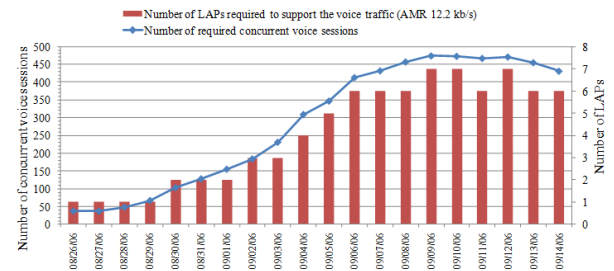

Figure 4. Evolution of the numbers of emergency voice traffic and required LAPs to support the subsequent concurrent voice sessions (AMR codec at a rate of 12.2 kb/s with a sample period of 20 ms and a MOS of 3.8).

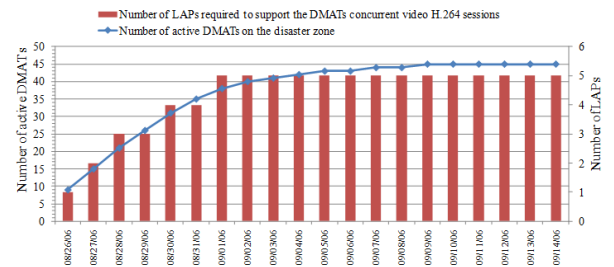

Figure 5. Evolution of the numbers of active DMATs and required LAPs to support the subsequent H.264 video traffic.

The different payload capabilities of the HAPs and LAPs will determine the number of modules that can be carried by these platforms and eventually the diversity of the services that can be provisioned from the ATNs. Moreover, platform station keeping, hand-off considerations even for fixed stations due to the platform movement and payload power bring lot of emphasis on the reliability of the aerial platforms in providing effective communications to the terrestrial stations.

The technological challenges and system issues that need to be addressed for effective integration of ATN platforms for emergency communications include the stratospherical environment and thermal conditions, launch and recovery, energy sources and platform aerodynamics. Apart from these, the technological choices for communication between the terrestrial responders and the aerial stations are an important aspect which governs the design of the system architecture, frequency planning, resource planning and network protocols. Moreover, the scale of disaster response requires flexible and incremental deployment strategies, including controlled mobility mechanisms and subsequently node coordination mechanisms with adequate positioning and clustering approaches to ensure maximum system availability and reliability.

Though energy efficiency is not the prime concern in a disaster scenario, it is of utmost importance to ensure the maximum operational lifetime for the ATN platforms and the terrestrial devices alike. This requires both efficient cooperative and context aware mechanisms which reduce the overhead in communications and techniques that scale down the power consumption in the uplink communications, apart from developing intelligent and energy efficient ground device interfaces. In addition, the ATN communications should complement well the established terrestrial disaster communication technologies like the TETRA wireless platforms and the COLT (Cells on Light Trucks) / COW (Cells on Wheels) platforms deployed by rescue operators [5]. The seamless integration of ATN with the satellite-terrestrial emergency service platforms and radical changes in the aeronautical and radio regulations can make integrated ATN a reality in providing emergency communications servicing large scale disasters.

## VI. CONCLUSION

In this paper, we discussed a novel integrated aerial architecture for emergency communications. We selected a subset of typical voice and video emergency services needed by rescuers, and estimated the subsequent impacts in terms of bandwidth requirements through the example of Hurricane Katrina. Estimation results showed that, in a case where traffic is ideally shared on the aerial platforms, a small number of LAPs have enough capacity to support a variable emergency traffic emitted on a large area. However, their limited coverage being an issue in large scale events, we also argued that the sustained performance of this infrastructure, notably in term of coverage, strongly relies on the ability to take advantage of the multi-level topology, including higher altitude platforms, and to enforce dedicated mechanisms, including controlled mobility.

In the future, we intend to study the impact of emergency traffic on the various communication interfaces of the considered architecture. In particular, we are interested in determining what proportion of the emergency traffic is meant to remain local, regional or to be transmitted through the backhaul links. We will also seek to further evaluate the interest of controlled mobility to support efficient cooperative communications between the different aerial platforms.